\documentclass[entropy,article,accept,moreauthors,pdftex]{Definitions/mdpi}
\usepackage{color}

\firstpage{1}
\pubvolume{24}
\issuenum{4}
\articlenumber{436}
\pubyear{2022}
\copyrightyear{2022}
\externaleditor{{Academic Editor: Anne Humeau-Heurtier}} 
\datereceived{7 February 2022}
\dateaccepted{18 March 2022}
\datepublished{22 March 2022}
\hreflink{https://doi.org/10.3390/e24040436} 

\Title{A Quantitative Comparison between Shannon and Tsallis--Havrda--Charvat Entropies Applied to Cancer Outcome~Prediction}

\TitleCitation{A Quantitative Comparison between Shannon and Tsallis--Havrda--Charvat Entropies Applied to Cancer Outcome Prediction}

\Author{Thibaud Brochet $^{1}$, Jérôme Lapuyade-Lahorgue $^{1}$, Pierre Vera $^{2}$ and Su Ruan $^{1,}$*\href{https://orcid.org/0000-0001-8785-6917}{\orcidicon}}

\AuthorNames{Thibaud Brochet, Jérôme Lapuyade-Lahorgue, Pierre Vera and Su Ruan}

\AuthorCitation{\textls[15]{Brochet, T.; Lapuyade-Lahorgue}, J.; \linebreak Vera, P.; Ruan, S.}

\address{$^{1}$ \quad LITIS, Quantif, University of Rouen, 76000 Rouen, France; thibaud.brochet@univ-rouen.fr (T.B.); jerome.lapuyade-lahorgue@univ-rouen.fr (J.L.-L.)\\

$^{2}$ \quad Centre Henri Becquerel, 76038 Rouen, France; pierre.vera@chb.unicancer.fr}

\corres{Correspondence: su.ruan@univ-rouen.fr}

\abstract{In this paper, we propose to quantitatively compare loss functions based on parameterized Tsallis--Havrda--Charvat entropy and classical Shannon entropy for the training of a deep network in the case of small datasets which are usually encountered in medical applications. Shannon cross-entropy is widely used as a loss function for most neural networks applied to the segmentation, classification and detection of images. Shannon entropy is a particular case of Tsallis--Havrda--Charvat entropy.  In this work, we compare these two entropies through a medical application for predicting recurrence in patients with head--neck and lung cancers after treatment. Based on both CT images and patient information, a multitask deep neural network is proposed to perform a recurrence prediction task using cross-entropy as a loss function and an image reconstruction task. Tsallis--Havrda--Charvat cross-entropy is a parameterized cross-entropy with the parameter $\alpha$. Shannon entropy is a particular case of Tsallis--Havrda--Charvat entropy for $\alpha=1$. The influence of this parameter on the  final prediction results is studied.
In this paper, the experiments are conducted on two datasets including in total $580$ patients, of whom $434$ suffered from head--neck cancers and $146$ from lung cancers. The results show that Tsallis--Havrda--Charvat entropy can achieve better performance in terms of prediction accuracy with some values of $\alpha$.}

\keyword{deep neural networks; Shannon entropy; Tsallis--Havrda--Charvat entropy; generalized entropies; recurrence prediction; head--neck cancer; lung cancer}

\begin{document}

\section{Introduction}
This paper is devoted to studying the loss function based on Tsallis--Havrda--Charvat entropy in deep neural networks \cite{art1bis} for the prediction of outcomes in lung and head--neck cancers. When used for categorical prediction, the loss function is generally a cross-entropy related to a given entropy function. Indeed, cross-entropy-based loss functions are appropriate for evaluating how a probability distribution is close to the Dirac distribution. In deep neural networks, the output is a probability for each class obtained from a softmax activation function, and the Dirac distribution concentrated on one class represents the ground truth. There are several ways to compare these distributions  \cite{chung1989,budzynski2007,serrurier2013}. Entropy-based metrics, such as divergences and cross-entropies, are the most common because they are the most appropriate way to sum up the informative content of a distribution, as explained in \cite{kullbackLeibler}. In Ref.~\cite{art1}, different entropies are presented. In classification and prediction, the cross-entropy is derived from an entropy measure and used as a loss function measuring the difference between the predicted probability and the real Dirac probability. To sum up, in most neural networks used for prediction, Shannon-related cross-entropy is the most common and widely used for segmentation \cite{seg2004}, classification \cite{classif2005}, or detection \cite {detect2020} and many other applications \cite{art2,art3,art2bis,art3bis}. The reason why Shannon is the most used is twofold: first, because Shannon was the first entropy in the domain of information theory, and secondly, because it is extensive in the sense that the entropy of a multivariate distribution whose margins are independent is the sum of the marginal entropies. This last property makes the calculation of Shannon entropy easy. In Ref.~\cite{basseville}, different ways of choosing an entropy and an associated divergence are detailed. Among them, Shannon entropy can be extended by replacing the logarithm by another function. Cross-entropies can be defined by replacing the counting measure (resp. Lebesgue measure for continuous case) by a Radon--Nykodim derivative between probability measures. Shannon's entropy can be generalized on other entropies such as Renyi \cite{artRenyi} and Tsallis--Havrda--Charvat \cite{roselin2014,artIRBM}. In this paper, we are interested in a particular generalization of Shannon cross-entropy: Tsallis--Havrda--Charvat cross-entropy \cite{art4}. This class of entropies has the particularity of being parametrized with one parameter $\alpha$ and we recover Shannon entropy when the value of the parameter equal to $1$. The relevance and possibilities of Tsallis--Havrda--Charvat in the medical field have been discussed before and this paper expands on the  Tsallis--Havrda--Charvat formula.

Tsallis--Havrda--Charvat entropy was introduced independently by Tsallis \cite{tsallis88} in the context of statistical physics and by Havrda and Charvat \cite{havrda67} in the context of information theory. Tsallis--Havrda--Charvat entropy has been used in publications in several fields, including medical imaging \cite{rev1,rev2}. Tsallis--Havrda--Charvat entropy has rarely been used in deep learning, especially because of the difficulties with interpreting the hyperparameters $\alpha$. However, there exist some scientific articles on this issue. In Ref.~\cite{ramezani2021}, Tsallis--Havrda--Charvat entropy is used to reduce the loss while classifying an image. In Ref.~\cite{art4}, the authors define Tsallis--Havrda--Charvat entropy in terms of axiomatization and propose a generalization based on this. In Ref.~\cite{karmershu2003}, the maximization of the entropy measure is studied for different classes of entropies such as Tsallis--Havrda--Charvat; the maximization of Tsallis--Havrda--Charvat entropy under constraints appears to be a way to generalize Gaussian distributions. In our previous work \cite{artIRBM}, Tsallis--Havrda--Charvat cross-entropy is used for the detection of noisy images in pulmonary microendoscopy. To capitalize and improve on this previous work, we propose to use deep learning in this paper. It allows for taking advantage of the previously used architecture by tuning and improving it to achieve better results. Deep learning is also very relevant for our field of study.

\begin{figure}[H]
\begin{minipage}[b]{.48\linewidth}
\centering
\hspace{-70pt} \centerline{\includegraphics[height=4.0cm,width=4.0cm]{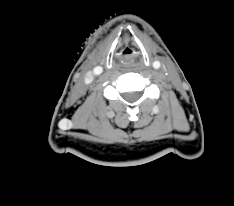}}
\vspace{0.2cm}

\end{minipage}
\hfill
\begin{minipage}[b]{0.48\linewidth}
\centering
\hspace{-70pt} \centerline{\includegraphics[height=4.0cm,width=4.0cm]{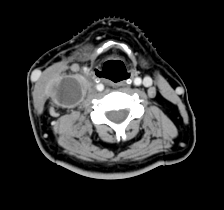}}
\vspace{0.2cm}

\end{minipage}
\hfill
\begin{minipage}[b]{.48\linewidth}
\centering
\hspace{-70pt} \centerline{\includegraphics[height=4.0cm,width=4.0cm]{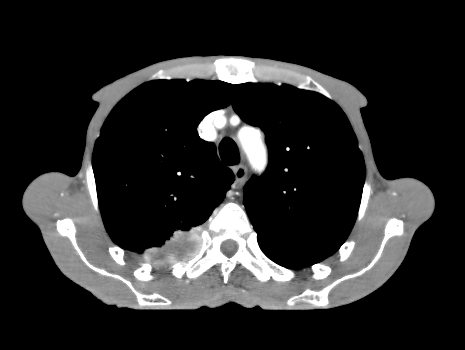}}
\vspace{0.2cm}

\end{minipage}
\hfill
\begin{minipage}[b]{0.48\linewidth}
\centering
\hspace{-70pt} \centerline{\includegraphics[height=4.0cm,width=4.0cm]{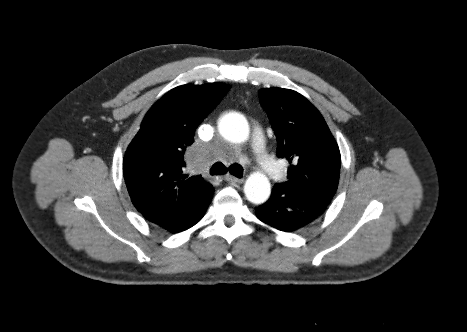}}
\vspace{0.2cm}

\end{minipage}

\caption{ Input images: head--neck CT (\textbf{above}) and lung CT (\textbf{below}).}
\label{fig:imgs}
\end{figure}

Deep learning has been widely developed in the medical field for classification or segmentation tasks \cite{zhou2021,amyar2020,art1ter}. Classification can be used to identify automatically the kind of cancer from which the patient is suffering \cite{art2ter,art3ter} or the relevant outcomes after treatment, such as survival expectation \cite{art4ter} or relation to the treatment \cite{amyar2019}. Recurrence in cancer after treatment is one of the main concerns for physicians \cite{art5ter}, as it can dramatically impact the outcome for patients and their life expectancy. It would be beneficial for treatment selection if one could predict whether a recurrence will occur. Some studies have been carried out using CT scan images and clinical data. To our knowledge, there is no article using Tsallis--Havrda--Charvat for recurrence prediction \cite{artIRBM}.
The novelty of this article lies in the performance comparison of Shannon and Tsallis--Havrda--Charvat entropies in the context of cancer recurrence prediction with CT scan data combined with clinical information for patients affected by head and neck (H\&N) or lung cancers (examples of the analyzed CT images are displayed in Figure \ref{fig:imgs}). Moreover, we decided to study the parameter value in particular to examine its impact in order to predict these recurrences in both kinds of cancer. As medical data are generally scarce, the choice of a good entropy is important even if it can improve the performance only by $1$ or $2$ percent.

The paper is organized as follows. In the first section, we recall how categorical Shannon cross-entropy is defined and how it can be generalized for Tsallis--Havrda--Charvat. The second section is devoted to experiments and a comparison between the two entropies.

\vspace{-9pt}
\section{Entropy}
As our problem concerns binary prediction, we focused only on finite-state random variables whose state-space was provided with the counting measure. Obviously, these results can be generalized for the finite-dimensional vectorial space $\mathbb{R}^{n}$ provided with the Lebesgue measure.
\subsection{Shannon Entropy and Related Cross-Entropy}
For a discrete random variable $Y$, taking its values in $\Omega=\{1,\ldots,k\}$ with respective probabilities $p_1,\ldots,p_k$, Shannon entropy is defined by:
\begin{equation}
\label{eq:shannon_entropy}
H(p)=-\sum_{i=1}^k\log(p_i)\times p_i
\end{equation}
Shannon is minimal and equal to $0$ if $p_i=1$ for one $i$ and $0$ otherwise; it is maximal if $Y$ is uniformly distributed. The corresponding cross-entropy is given by:
\begin{equation}
\label{eq:cross_entropy}
H(q:p)=-\sum_{i=1}^{k}\log(q_i)p_i
\end{equation}
Generally, in a classification problem, the true distribution $p$ is a Dirac distribution $\delta_{i_0}$, where $i_0$ is the class of the data. In this case, the cross-entropy is $H(p;q)=-\log(q_{i_0})\geq 0$. As a consequence, $q_{i_0}$ is closer to $0$ and $H(p;q)$ is higher. Furthermore, minimizing the cross-entropy forces $q$ to be as close as possible to the distribution $p$, and as this last one is the Dirac distribution $\delta_{i_0}$, the minimum of the cross-entropy is $0$.
\subsection{Tsallis--Havrda--Charvat Cross-Entropy}
There are several ways to generalize Shannon entropy, as explained in \cite{basseville}. Shannon entropy can be expressed by:
\begin{equation}
H(q)=-\sum_{i=1}^{k}h(q_i)
\end{equation}
where $h(u)=u\log(u)$. $h$ is a convex function such that $h(1)=0$. The idea is to choose another function satisfying the same properties. Tsallis--Havrda--Charvat entropy is defined by choosing:
\begin{equation}
h_{\alpha}(u)=\frac{u^{\alpha}-u}{\alpha-1},
\end{equation}
where $\alpha>0$ and is given by:
\begin{equation}
\label{eq:havrda_charvat}
H_{\alpha}(q)=\frac{1}{\alpha-1}\times\left[1-\sum_{i=1}^{k}q_i^{\alpha}\right]
\end{equation}
The associated cross-entropy is given by:
\begin{equation}
\label{eq:cross_havrda_charvat}
H_{\alpha}(q:p)=\frac{1}{\alpha-1}\times\left[1-\sum_{i=1}^{k}q_i^{\alpha-1}p_i\right]
\end{equation}
As for classical cross-entropy, Tsallis--Havrda--Charvat cross-entropy forces the predicted distribution $q$ to be as close as possible to $p$ when $p$ is a Dirac distribution.

\section{Neural Network Architecture for Recurrence Prediction}
\begin{figure}[H]
\includegraphics[width=10.5 cm]{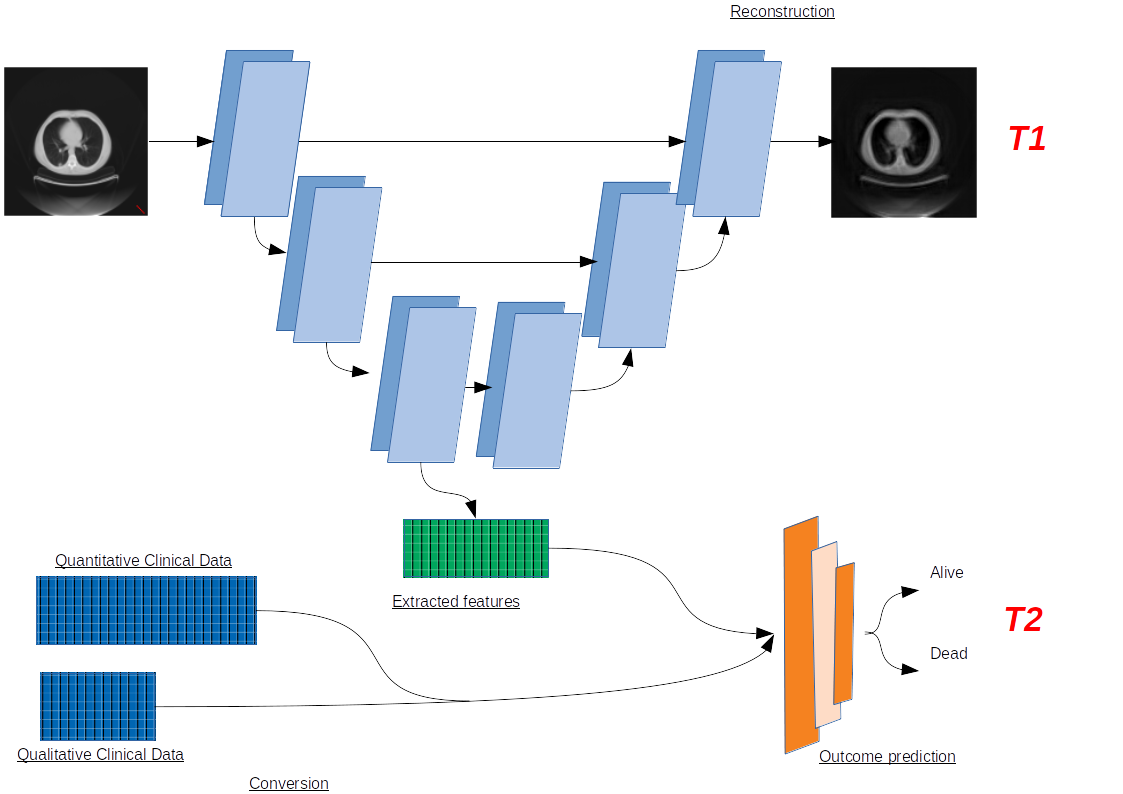}
\caption{ Architecture of the multitask neural network for recurrence prediction (T2) with the help of another task (T1: image reconstruction).\label{fig:arch}}
\end{figure}

The proposed architecture used for recurrence prediction is a multitask neural network with a U-Net backbone, with one branch able to perform prediction tasks and another to reconstruct the input image for extracting features to help prediction.
The architecture is presented in Figure \ref{fig:arch}.
The U-Net backbone is composed of three convolutional layers skipped by concatenations to add the features extracted from descending convolutions to the ascending ones.
Within the convolution layers, we use ReLU activation functions.

At the bottom of the network, the extracted features are sent as inputs to one branch of fully connected layers in charge of making a decision path to determine whether the patient is at risk of recurrence.

Two main tasks are jointly carried out by the network. T1 is the reconstruction task, specific to the U-Net part of the network. It allows for determining whether the features extracted by the descending part of the U are relevant for prediction and classification and representative of the whole CT scan at the same time. The loss function used in this task is the mean squared error. It is defined as follows:
\begin{equation}
\label{eq:MSE}
L_{\textrm{rec}}=\frac{1}{N}\sum_{n=1}^{N}\Vert \vec{\textrm{\textbf{y}}}_n-\hat{\vec{\textrm{\textbf{y}}}}_n\Vert^{2},
\end{equation}
where $\vec{\textrm{\textbf{y}}}_n$ represents the true data values for the $n$-th patient and $\hat{\vec{\textrm{\textbf{y}}}}_n$ is the estimated output from the network, with $\Vert .\Vert$ being the Euclidean norm and \emph{N} the number of patients.

The mean squared error computes the squared distance between the predicted output and the input image volumes. This function allows for comparing the predicted images and the true ones voxel-by-voxel and to train the network to recompose the images from the extracted features.

T2 is the prediction task. It is used to determine, from the same input, whether the current patient is at risk of encountering a recurrence of their cancer. It is constructed with fully connected layers and ends up in a binary classification.
The prediction task's loss function was the subject of our tests. We compared two entropies through this task.

The first, and most commonly used, was Shannon's binary cross-entropy.
\begin{equation}
\label{eq:BCE}
L_{\textrm{pred},1}=-\frac{1}{N}\sum_{n=1}^{N}\left[p_n \times \log(q_n) + (1-p_n) \times \log(1-q_n)\right],
\end{equation}
where $p_n$ is the true class, $p_n=1$ if recurrence, $p_n=0$ otherwise, and $q_n$ is the estimated probability of recurrence, with \emph{N} being the number of patients.

The second entropy was the generalized formula, Tsallis--Havrda--Charvat binary cross-entropy.

\begin{equation}
\label{eq:HCDCE}
L_{\textrm{pred},\alpha}=\frac{1}{\alpha-1}\times\left[1-\frac{1}{N}\sum_{n=1}^{N}(q_n^{\alpha-1}p_n +(1-q_n)^{\alpha-1}(1-p_n))\right]
\end{equation}

Binary cross-entropies are loss functions that are able to compare binary predictions with ground truths, which makes them relevant for our binary labels.

The total loss function of the network is the sum of the two losses.
The choice of this total loss function was motivated by different experiments in which we used different weights for the individual loss functions. It appears that the sum with equal weights provided the best results.
\begin{equation}
\label{eq:TOT}
L_{total}=L_{\textrm{rec}}  + L_{\textrm{pred},\alpha}
\end{equation}

The prediction branch is the subject of interest for this article, with the reconstruction task being used to help the prediction in the feature extraction step.

Regarding the execution time and its change owing to the increasing or decreasing complexity of the network, we proposed to lock our network at a certain level of complexity and find an acceptable execution time of 5--7 h for a number of epochs of 100. This was considered to be a good compromise between obtaining a sufficient number of epochs to achieve meaningful results and not obtaining so many as to cause overfitting. We decided to use convolution layers in our U-Net backbone because of the limited computational power of available machines and thus the need to limit the number of parameters in the network.
Allowing for a few hours of computations enabled the experiments to be run at night so that the results could be available and ready for analysis the next day.

\section{Experiments}

\subsection{Datasets}

The datasets were composed of 580 patients, among which 434 suffered from head--neck cancer and 146 from lung cancer. Both datasets were small in size. We chose to conduct experiments on the two subsets (head--neck and lung) separately. Indeed, the optimal value of $\alpha$ depends on the kind of data used. Moreover, we had already tested the total dataset of 580 and the results were poor. We therefore chose to show only the results for separate~datasets.

CT images used in the neural network were resized with an image resolution of \mbox{128 $\times$ 128 $\times$ 64 voxels.} The patient information used as input data in the neural network was of two kinds, namely quantitative and qualitative, as shown in Tables \ref{tab:clinN} and \ref{tab:clinL}.

\begin{table}[H]
\small
\caption{Quantitative clinical data processed through the network.}\label{tab:clinN}
\setlength{\tabcolsep}{17.15mm}
\begin{tabular}{cc}
\toprule
\textbf{ Clinical Data} & \textbf{Modality}\\
\midrule
Hemoglobin &g/dL \\
Lymphocytes& Giga/L \\
Leucocytes& Giga/L\\
Thrombocytes &Giga/L \\
Albumin &g/L\\
Treatment duration &Days\\
Total irradiation dose &Gy\\
Number of fractions& / \\
Average dose per fraction& Gy\\
Weight at the start and end of treatment &kg\\
\bottomrule
\end{tabular}
\end{table}

\vspace{-10pt}
\begin{table}[H]
\small
\caption{Qualitative clinical data processed through the network.} \label{tab:clinL}
\setlength{\tabcolsep}{9.95mm}
\begin{tabular}{cc}
\toprule
\textbf{Clinical Data} & \textbf{Modality}\\
\midrule
Gender& M/F\\
Tabacology &Smoker, non-smoker, former smoker\\
Use of induction chemotherapy &Yes/No\\
Use of concomitant chemotherapy &Yes/No\\
TNM & Tumor, Node, Metastasis\\
\bottomrule
\end{tabular}
\end{table}

\vspace{-2pt}
Our experiments consisted of comparing the accuracy of Tsallis--Havrda--Charvat and Shannon for both datasets.

\subsection{Evaluation Method}

Since the study was conducted on small datasets (434 and 146 patients), a result validation strategy was required. We proposed to use the k-fold cross-validation relevant for small data validation. In our work, we used a five-fold cross-validation.

The procedure unfolds as follows:
\begin{itemize}
\item{Shuffle the dataset;}
\item{Split it randomly into 5 subsets;}
\item{For each subset:}
\subitem{Take the subset as a test dataset;}
\subitem{Take the other sets as training data;}
\subitem{Fit the model to the training data and evaluate it on the test dataset;}
\subitem{Retain the evaluation score;}
\item{Summarize the skill of the model from the samples of model evaluation scores.}
\end{itemize}

Furthermore, accuracy was proposed as a metric for evaluation. It consisted, in this case, in comparing the values of the ground truth and the prediction and summing all these occurrences over the size of the dataset.

\subsection{Results}
\begin{figure}[H]
\begin{minipage}[b]{.48\linewidth}
\centering
\hspace{-70pt}\centerline{\includegraphics[height=4.0cm,width=4.0cm]{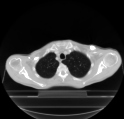}}
\vspace{0.2cm}

\end{minipage}
\hfill
\begin{minipage}[b]{0.48\linewidth}
\centering
\hspace{-70pt} \centerline{\includegraphics[height=4.0cm,width=4.0cm]{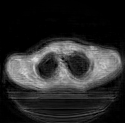}}
\vspace{0.2cm}

\end{minipage}
\hfill
\begin{minipage}[b]{.48\linewidth}
\centering
\hspace{-70pt}  \centerline{\includegraphics[height=4.0cm,width=4.0cm]{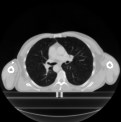}}
\vspace{0.2cm}

\end{minipage}
\hfill
\begin{minipage}[b]{0.48\linewidth}
\centering
\hspace{-70pt} \centerline{\includegraphics[height=4.0cm,width=4.0cm]{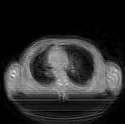}}
\vspace{0.2cm}

\end{minipage}

\centering
\caption{ Input images: original images (\textbf{left}) vs. reconstructed images (\textbf{right}).}
\label{fig:recons}
\end{figure}

The results achieved during the tests are displayed in this section. Reconstructed images are mainly used in order to show the relevance of extracted features for the prediction. Therefore, their performance is not very important here, because the objective is the prediction of recurrence. The original images and reconstructed ones are featured in \mbox{Figure \ref{fig:recons}.} We can see that the reconstructed images are similar to the input images, meaning that our network is able to recover input images.

The loss of information generates uncertainty in each image. A possible improvement would be to use a fuzzy image processor to improve the quality of the obtained images, as described in \cite{doi443}.

\subsubsection*{Comparison Results}

Regarding Tsallis--Havrda--Charvat, we studied its hyperparameter $\alpha$ as varying from 0.1 to 2.0. When $\alpha=1$, this entropy corresponds to Shannon entropy. The displayed \emph{p}-value measures whether the results acquired by Tsallis--Havrda--Charvat five-fold cross-validation are statistically different from Shannon's. Two conditions must be satisfied in order to accept the Tsallis--Havrda--Charvat entropy as providing better results than Shannon entropy: the average of the five-fold results must be superior to Shannon and the \emph{p}-value must be smaller than $0.05$. The results are described in the following tables.

The results achieved for the dataset of head--neck cancers are described in Table \ref{tab:2bis}.

Regarding the dataset containing lung cancers, the results achieved are described in Table \ref{tab:2ter}.

After fine-tuning, we obtained a set of optimal hyperparameters.

The results achieved via the Tsallis--Havrda--Charvat formula confirm that, for most values of the hyperparameter $\alpha$, the final accuracy is not superior to the accuracy obtained by Shannon's loss function. It can also be observed that the loss function derived from Havrda--Charvat equation can provide better results than Shannon's in some cases. However, it is difficult to know a priori what value of $\alpha$ is good for an application. Its choice is still a challenge.

We highlighted the results providing both better accuracy and a significant \emph{p}-value in blue. The most promising values were achieved for $\alpha$ equal to 1.5 with head--neck cancers and between 1.9 and 3.5 with lung cancer. In this regard, we can state that the results obtained by Tsallis--Havrda--Charvat can be significantly improved compared to Shannon.

When analyzing the lung cancer results, a plateau can be easily noticed where Tsallis--Havrda--Charvat achieves better results than Shannon. This involves a specific set of values of $\alpha$, from 1.9 to 3.5, in which the Tsallis--Havrda--Charvat loss function is significantly more efficient than Shannon's.

\begin{table}[H]
\small
\caption{Accuracy obtained by loss function derived from Tsallis--Havrda--Charvat entropy in a function of $\alpha$ for the head--neck cancer dataset (\emph{p}-values lower than 0.05 and accuracies higher than Shannon's are highlighted in blue).
\label{tab:2bis}}
\setlength{\tabcolsep}{3.00mm}
\begin{tabular}{ccccccccc}
\toprule
& \multicolumn{5}{c}{\textbf{5-Fold}} & & & \\
\boldmath{$\alpha$ } & \textbf{1}& \textbf{2} & \textbf{3} & \textbf{4} &\textbf{5} & \textbf{Average} & \textbf{SD}& \textbf{\emph{p}-Value}\\
\midrule
0.1 &  0.68&0.53&0.6&0.58&0.63&0.60&0.06&0.01\\

0.3 &  0.60&0.70&0.70&0.70&0.43&0.63&0.12&0.28\\

0.5 & 0.58&0.58&0.60&0.70&0.73&0.64&0.07&0.27\\

0.7 &0.85&0.70&0.60&0.70&0.65&0.70&0.09&0.25\\

0.9 & 0.58&0.60&0.60&0.60&0.68&0.61&0.04&0.07\\
\textcolor{red}{1.0} &  \textcolor{red}{0.75}&\textcolor{red}{0.65}  &\textcolor{red}{0.68} &\textcolor{red}{0.58} &\textcolor{red}{0.70} &\textcolor{red}{0.67} &\textcolor{red}{0.06}&\textcolor{red}{N/A (Shannon entropy)}\\
1.1 & 0.68&0.75&0.75&0.73&0.75&0.73&0.03&0.09\\

1.3 &  0.63&0.73&0.68&0.75&0.70&0.70&0.05&0.32\\

\textcolor{blue}{1.5} &\textcolor{blue}{ 0.70}&\textcolor{blue}{0.78}&\textcolor{blue}{0.85}&\textcolor{blue}{0.80}&\textcolor{blue}{0.88}&\textcolor{blue}{0.80}&\textcolor{blue}{0.07}&\textcolor{blue}{0.03}\\

1.7 &0.80&0.73&0.63&0.75&078&0.74&0.07&0.07\\

\textcolor{blue}{1.9} & \textcolor{blue}{ 0.75}& \textcolor{blue}{0.73}& \textcolor{blue}{0.73}& \textcolor{blue}{0.75}& \textcolor{blue}{0.83}& \textcolor{blue}{0.76}& \textcolor{blue}{0.05}& \textcolor{blue}{0.02}\\

2.1&0.68&0.63&0.60&0.62&0.575&0.63&0.04&0.12\\

2.3&0.73&0.73&0.73&0.73&0.7&0.72&0.01&0.09\\

2.5&0.75&0.58&0.68&0.68&0.6&0.66&0.07&0.34\\

2.7&0.68&0.53&0.45&0.63&0.6&0.58&0.09&0.04\\

2.9&0.73&0.73&0.73&0.75&0.73&0.73&0.01&0.07\\

3.1&0.7&0.55&0.65&0.57&0.63&0.62&0.06&0.02\\

3.3&0.65&0.65&0.65&0.58&0.55&0.62&0.05&0.07\\

3.5&0.73&0.75&0.73&0.75&0.70&0.73&0.02&0.08\\

3.7&0.73&0.70&0.60&0.60&0.58&0.64&0.07&0.20\\

3.9&0.68&0.70&0.55&0.58&0.53&0.61&0.08&0.09\\

\bottomrule
\end{tabular}
\end{table}
\vspace{-12pt}

\begin{table}[H]
\small
\caption{Accuracy obtained by loss function derived from Tsallis--Havrda--Charvat entropy in a function of $\alpha$ for the lung cancer dataset (\emph{p}-values lower than 0.05 and accuracies higher than Shannon's are highlighted in blue).
\label{tab:2ter}}
\setlength{\tabcolsep}{3.00mm}
\begin{tabular}{ccccccccc}
\toprule
& \multicolumn{5}{c}{\textbf{5-Fold}} & & & \\
\textbf{$\alpha$ } & \textbf{1}& \textbf{2} & \textbf{3} & \textbf{4} &\textbf{5} & \textbf{Average} & \textbf{SD}& \textbf{\emph{p}-Value}\\
\midrule
0.1 &  0.58&0.58&0.47&0.58&0.63&0.57&0.06&0.23\\

0.3 & 0.58&0.58&0.58&0.58&0.68&0.60&0.04&0.13\\

0.5 & 0.63&0.58&0.52&0.58&0.52&0.56&0.03&0.26\\

0.7 &0.58&0.58&0.63&0.63&0.58&0.60&0.03&0.13\\

0.9 & 0.63&0.53&0.68&0.47&0.53&057&0.08&0.21\\
\textcolor{red}{1.0} &  \textcolor{red}{0.73}&\textcolor{red}{0.47}  &\textcolor{red}{0.47} &\textcolor{red}{0.47} &\textcolor{red}{0.47} &\textcolor{red}{0.52} &\textcolor{red}{0.12}&\textcolor{red}{N/A (Shannon entropy)}\\
1.1 & 0.63&0.63&0.52&0.52&0.52&0.56&0.06&0.18\\

1.3 & 0.68&0.53&0.53&0.47&0.53&0.55&0.08&0.15\\

1.5 & 0.58&0.53&0.53&0.47&0.53&0.53&0.04&0.44\\

1.7 &0.73&0.47&0.63&0.57&0.42&0.56&0.12&0.37\\

\textcolor{blue}{1.9} & \textcolor{blue}{0.69}&\textcolor{blue}{0.63}&\textcolor{blue}{0.58}&\textcolor{blue}{0.53}&\textcolor{blue}{0.68}&\textcolor{blue}{0.62}&\textcolor{blue}{0.07}&\textcolor{blue}{0.04}\\

\textcolor{blue}{2.1}& \textcolor{blue}{0.79}&\textcolor{blue}{0..84}&\textcolor{blue}{0.73}&\textcolor{blue}{0..79}&\textcolor{blue}{0.79}&\textcolor{blue}{0.79}&\textcolor{blue}{0.04}&\textcolor{blue}{0.004}\\

\textcolor{blue}{2.3}& \textcolor{blue}{0.84}&\textcolor{blue}{0.78}&\textcolor{blue}{0.84}&\textcolor{blue}{0.73}&\textcolor{blue}{0.84}&\textcolor{blue}{0.81}&\textcolor{blue}{0.05}&\textcolor{blue}{0.002}\\

\textcolor{blue}{2.5}& \textcolor{blue}{0.79}&\textcolor{blue}{0.84}&\textcolor{blue}{0.74}&\textcolor{blue}{0.68}&\textcolor{blue}{0.63}&\textcolor{blue}{0.74}&\textcolor{blue}{0.08}&\textcolor{blue}{0.007}\\

\textcolor{blue}{2.7}& \textcolor{blue}{0.79}&\textcolor{blue}{0.74}&\textcolor{blue}{0.69}&\textcolor{blue}{0.74}&\textcolor{blue}{0.74}&\textcolor{blue}{0.74}&\textcolor{blue}{0.04}&\textcolor{blue}{0.003}\\

\textcolor{blue}{2.9}& \textcolor{blue}{0.79}&\textcolor{blue}{0.79}&\textcolor{blue}{0.79}&\textcolor{blue}{0.79}&\textcolor{blue}{0.73}&\textcolor{blue}{0.78}&\textcolor{blue}{0.03}&\textcolor{blue}{0.004}\\

\textcolor{blue}{3.1}& \textcolor{blue}{0.74}&\textcolor{blue}{0.74}&\textcolor{blue}{0.78}&\textcolor{blue}{0.78}&\textcolor{blue}{0.68}&\textcolor{blue}{0.74}&\textcolor{blue}{0.04}&\textcolor{blue}{0.008}\\

\textcolor{blue}{3.3}& \textcolor{blue}{0.79}&\textcolor{blue}{0.79}&\textcolor{blue}{0.74}&\textcolor{blue}{0.74}&\textcolor{blue}{0.74}&\textcolor{blue}{0.76}&\textcolor{blue}{0.03}&\textcolor{blue}{0.003}\\

3.5&0.74&0.73&0.68&0.33&0.58&0.61&0.17&0.12\\

3.7&0.68&0.63&0.47&0.63&0.73&0.63&0.09&0.07\\

3.9&0.78&0.53&0.47&0.47&0.63&0.58&0.13&0.07\\
\bottomrule
\end{tabular}
\end{table}

\section{Discussion}

It has been determined that the Tsallis--Havrda--Charvat loss function performs equally well or better than Shannon cross-entropy, depending on the value of its hyperparameter $\alpha$.\\
It can be said that the Tsallis--Havrda--Charvat loss function, depending on the value of its hyperparameter $\alpha$, can fit a wider array of input data and can potentially yield better~results.

Conversely, we can state that, based on the calculated p-values and standard deviations, Tsallis--Havrda--Charvat entropy seems more unstable than Shannon entropy, as its standard deviation may reach 0.12, where Shannon's is only 0.06. Furthermore, when looking at the p-values for several values of $\alpha$, the results of Havrda--Charvat are not statistically different from Shannon's. The instability of the results obtained using Tsallis--Havrda--Charvat could be explained by the fact that, despite the five-fold method, the data are still too scarce to reach a stable answer. In addition, data are tridimensional, making it more difficult than with 2D images to extract relevant features and thereby complicating the network's tasks even further. In the  analysis of 3D images, multiple slices must be taken into account in order to make a decision, which drastically increases the number of variables to be learned by the neural network. Moreover, in order to be usable for analysis by the neural network, the data are all supposed to have the same size. This is why, as for reasons of computational power, it was decided to have the data resized to \mbox{$128\times 128\times 64$ voxels.} This size, despite still being full of information for the network, implies a loss of information for wider, larger and deeper images.

Nevertheless, the value of $\alpha$ plays a large part in the behavior of the loss function, and it is the key element that can be set to fit the input data, but the questions remains: what $\alpha$ fits which data?
The choice of the value of the hyperparameter $\alpha$ remains a challenge as it depends strongly on the kind of data. In perspective, it would be interesting to develop an algorithm for selecting automatically the value of this hyperparameter in order to fit it as accurately as possible to the data.
The aim is to reach the plateau, or area, of $\alpha$ where the Tsallis--Havrda--Charvat loss function provides consistently better results and features a smaller SD and \emph{p}-value. A further analysis needs to be conducted on the link between input data and the location of the best $\alpha$ area in order to determine the kind of extracted feature and the kind of neuronal path that cause one area to be more efficient than another. For instance, in our case, the question is about which key feature of the input lung cancer images leads to better values between 1.9 and 3.5 and which key feature of the input H\&N cancer images lead to better results for $\alpha$ equal to 1.5.

\section{Conclusions}

In this article, we established that, for our data and in some cases, Tsallis--Havrda--Charvat cross-entropy performs better than a Shannon-based loss function. Tsallis--Havrda--Charvat performed best on the head--neck dataset and lung dataset, at 80\% and 81\%, respectively, of correct recurrence prediction, while Shannon's results for these two datasets were 67\% and 52\%, respectively. This makes the Tsallis--Havrda--Charvat formula the best candidate for further research on these datasets. For further research we might adapt Tsallis--Havrda--Charvat binary cross-entropy to a categorical cross-entropy. This would allow for making multi-class predictions, including estimating the time between the end of cancer treatment and recurrence. Another axis of evolution could be finding a way to automatically determine the proper value of $\alpha$ for a given application.


\vspace{6pt}

\authorcontributions{Conceptualization, S.R. and J.L.-L.; methodology, S.R., T.B. and J.L.-L.; medical dataset and medical expertise, P.V.; software, T.B. and J.L.-L.; validation, S.R.; formal analysis, T.B.; investigation, S.R., T.B. and J.L.-L.; writing---original draft preparation, T.B.; writing---review and editing, J.L.-L. and S.R.; supervision, S.R.; project administration, S.R. All authors have read and agreed to the published version of the manuscript.
}
\funding{This project was co-financed by the European Union with the European regional development fund (ERDF, 18P03390/18E01750/18P02733), by the Haute-Normandie Regional Council via the M2SINUM project and by the PRPO project from the Cancéropôle Nord-Ouest, France.}

\institutionalreview{Not applicable.}

\informedconsent{Not applicable.}

\dataavailability{The data are not publicly available due to being the property of Centre Henri Becquerel.}

\conflictsofinterest{The authors declare no conflict of interest.}



\abbreviations{Abbreviations}{
The following abbreviations are used in this manuscript:\\

\noindent
\begin{tabular}{@{}ll}
MDPI & Multidisciplinary Digital Publishing Institute\\
DOAJ & Directory of open access journals\\
TLA & Three-letter acronym\\
LD & Linear dichroism\\
H\&N & Head--neck\\
H-Cd & Havrda--Charvat\\
Gy & Gray
\end{tabular}}



\begin{adjustwidth}{-\extralength}{0cm}

\reftitle{References}

\end{adjustwidth}

\end{document}